\documentclass[twocolumn,preprintnumbers,nofootinbib]{revtex4-1}
\usepackage{amssymb}
\usepackage{graphicx}
\usepackage{epsfig}
\usepackage{psfrag}

\usepackage{color}
\usepackage{enumerate}
\def\beq{\begin{equation}}
\def\eeq{\end{equation}}
\def\beqn{\begin{eqnarray}}
\def\eeqn{\end{eqnarray}}

\renewcommand{\texttt}{{}}
\newcommand{\be}{\begin{eqnarray}}
\newcommand{\ee}{\end{eqnarray}}

\begin{document}


\title{Terminating black holes in asymptotically free quantum gravity}

\author{Cosimo Bambi}
\email{bambi@fudan.edu.cn}

\author{Daniele Malafarina}
\email{daniele@fudan.edu.cn}

\author{Leonardo Modesto}
\email{lmodesto@fudan.edu.cn}

\affiliation{Center for Field Theory and Particle Physics \& Department of Physics, 
Fudan University, 200433 Shanghai, China}

\date{\today}

\begin{abstract}
We study the homogeneous gravitational collapse of a spherical cloud of matter 
in a super-renormalizable and asymptotically free theory of gravity. 
We find a picture that differs substantially from the classical scenario. 
The central singularity appearing in classical 
general relativity is replaced by a bounce, after which the cloud re-expands 
indefinitely. We argue that a black hole, strictly speaking, never forms. The collapse 
only generates a temporary trapped surface, which can be interpreted as a black 
hole when the observational timescale is much shorter than the one of the collapse. 
However, it may also be possible that the gravitational collapse produces a black hole 
and that after the bounce the original cloud of matter evolves into a new universe.
\end{abstract}

\pacs{04.60.Bc, 04.70.Dy, 95.30.Sf}

\maketitle


\section{Introduction}

When a star exhausts all its nuclear fuel, the thermal pressure of its particles
cannot compensate the gravitational force any more, and the body contracts 
until it finds a new equilibrium configuration. For very massive stars, there is no 
known mechanism capable of compensating their own gravitational force, 
and the body will undergo a complete gravitational collapse. In general 
relativity, under the assumptions of the validity of the strong energy 
condition and of the existence of global hyperbolicity, the final product of the 
collapse is a singularity of the spacetime~\cite{sing,sing2}. At the singularity, 
predictability is lost and standard physics breaks down. According to the weak 
cosmic censorship conjecture, singularities produced in the gravitational 
collapse must be hidden behind an event horizon and the final product of the 
collapse is a black hole~\cite{wccc}. The energy conditions and the cosmic 
censorship conjecture are two fundamental ingredients in the theory of black 
hole physics, as they regulate most of the properties of their horizons (see 
e.g. Ref~\cite{sing2}, chapter~9).

In this paper, we study the gravitational collapse in a large class of non-local 
theories of gravity, which also includes models inspired by string field 
theory~\cite{string}. We consider the simplest cases of homogeneous 
collapse of a spherical cloud of dust and radiation and we find a new picture
for the gravitational collapse. The spacetime singularity appearing in classical 
general relativity is replaced by a bounce, after which the cloud re-expands 
indefinitely. It seems that black holes, strictly speaking, never form, in the sense 
that there are no regions causally disconnected from future null infinity. We find 
that the collapse produces a temporary trapped surface, 
which appears like the classical apparent horizon in the weak field regime and 
is removed when size and density lead to the regime of asymptotic freedom.
The object can look like a black hole to far away observers when the 
observational timescale is much shorter than the one of 
the collapse. In terms of the effective theory, in which the Einstein
equations are satisfied by an effective perfect fluid 
matching at the boundary to a Vaidya solution, the disappearance 
of the horizon can be seen as the result
of an ingoing flux of negative energy. 
One practical consequence of the model is that astrophysical black hole candidates 
should be characterized by an apparent mass loss, which might produce 
some observational effects. While we study in some detail only a specific 
model, we argue that this picture for the gravitational collapse must be 
common to many quantum gravity theories. Independently of the exact UV 
completion, in several models quantum corrections make gravity repulsive at 
very high densities~\cite{frolov}, and this is the key-ingredient to get our result.

\section{Theoretical framework}

As classical action, we consider a ``non-polynomial" or ``semi-polynomial" 
extension of Stelle's quadratic theory~\cite{stelle}, in which the dimensionless 
coupling constants are replaced by an entire function of the D'Alembertian 
operator~\cite{sn,efimov}  
\be
\hspace{-0.2cm}
\mathcal{S} = \int d^4 x \frac{2 \sqrt{|g|}}{\kappa^2} 
\Big[R - G_{\mu\nu} \frac{ V(-\Box/\Lambda^2)^{-1} -1}{\Box} 
R^{\mu\nu} \Big] \, ,
\label{theory}
\ee
where $G_{\mu\nu}$ is the Einstein tensor and $\kappa^2 = 32 \pi G_{\rm N}$. 
All the non-polynomiality is in the form 
factor $V(-\Box/\Lambda^2)$, where $\Lambda$ is the Lorentz invariant energy scale. 
$\Lambda$ is not subject to infinite or finite (non analytic) renormalizations, and 
it is only constrained to be large by observations. The natural value of
$\Lambda$ is of order the Planck mass. At the classical level, all the corrections to 
the Einstein-Hilbert action are suppressed by $1/\Lambda$, and therefore the 
theory reduces to general relativity at low energies. That is also true at the quantum level, 
as a consequence of the Donoghue argument~\cite{dono} (see Appendix~\ref{s-a-1} 
for more details).

The entire function $V(-\Box/\Lambda^2)$ must have no poles in the whole 
complex plane, in order to ensure unitarity, and must exhibit at least logarithmic 
behavior in the ultraviolet regime, to give super-renormalizablitity at the quantum 
level. General form factors suggested by a class of theories consistent at the quantum 
level are
\be
&&
V(z)^{-1} 
= \left| p_{\gamma + 1}(z) \right|  \, e^{\frac{1}{2} \left[ 
\Gamma \left(0, p_{\gamma + 1}^2(z) \right)+\gamma_E  \right] } \, , 
\label{Tombola}\\
&& V(z)^{-1}
= e^{z^n} \,\,\,\, n \in \mathbb{N}^{+},
\label{VK}
\ee
where $z \equiv - \Box/\Lambda^2$ and $p_{\gamma + 1}(z)$ is a real polynomial 
of degree $\gamma + 1$ ($\gamma \in \mathbb{N}$, $\gamma >2$). The theory
is uniquely specified once the form factor is fixed, because the latter does not 
receive any renormalization: the ultraviolet theory is dominated by the bare action 
(that is, counterterms are negligible). In this class of theories, we only have the 
graviton pole. Since $V(z)$ is an entire function, there are no ghosts and no 
tachyons, independently of the number of time derivatives present in the action.

Concerning the difficulties with particular form factors and non-local operators,
we note that the class of operators introduced by Krasnikov -- $V(z)^{-1}$ given
by Eq.~(\ref{VK}) with $n$ even -- and the one introduced by Tomboulis --
$V(z)^{-1}$ given by Eq.~(\ref{Tombola}) -- are well defined in the 
Euclidean as well as in the Lorentzian case, because $(k_{\rm E}^2)^2 = (k^2)^2$, 
where $k_{\rm E}$ is the momentum in the Euclidean space (see the paper by 
Krasnikov in Ref.~\cite{sn}). In what follows, we concentrate on the case 
in Eq.~(\ref{VK}) with $n=1$, which is suggested by string 
theory~\cite{string,afree}, but the qualitative behavior holds for a 
large class of models. In this paper we do all the calculations in the Lorentzian 
case, where the integral of interest is converging. To stress the genericity
of the result, independently of the Wick rotation, we present the solution 
with $n = 2$ in Appendix~\ref{s-a-3} and we have checked the solutions for 
other even values of $n$. There are no qualitative differences for different values
of $n$ in the physical quantities studied in this work.

\section{Homogeneous collapse}

The most general spherically symmetric metric describing a collapsing cloud 
of matter in comoving coordinates is given by
\begin{equation}\label{eq1}
    ds^2=-e^{2\nu}dt^2+\frac{R'^2}{G}dr^2+R^2d\Omega^2 \; ,
\end{equation}
where $d\Omega^2$ represents the line element on the unit two-sphere and 
$\nu$, $R$, and $G$ are functions of $t$ and $r$. In the homogeneous marginally 
bound case, we can choose $\nu = 0$ and $G=1$ (see e.g. Ref.~\cite{bmm}). 
The standard Einstein equations for the collapse of a perfect fluid are
\begin{equation}
\frac{\kappa^2}{4} \rho = \frac{F'}{R^2R'} \, , \qquad
\frac{\kappa^2}{4} p = -\frac{\dot{F}}{R^2\dot{R}} \, , \label{einstein-1}
\end{equation}
where the $'$ denotes a derivative with respect to $r$, and the $\dot{}$ denotes 
a derivative with respect to $t$. Here $\rho$ and $p$ are, respectively, the density and 
the pressure of the fluid, while $F$ is the Misner-Sharp mass, which is defined by 
$F = R \dot{R}^2$ and turns out to be twice the total gravitational mass contained 
within the shell labelled by $r$ at the time $t$. In the case of collapse, the usual 
prescription is that the area radius $R(r,t)$ is set equal to the comoving radius 
$r$ at the initial time $t_{\rm i}=0$, $R(r,0)=r$. We can then introduce a scale 
factor $a(t)$, $R(r,t)=r a(t)$, with $a(0) = 1$.

\begin{figure*}
\hspace{-0.3cm}
\includegraphics[type=pdf,ext=.pdf,read=.pdf,width=5.43cm]{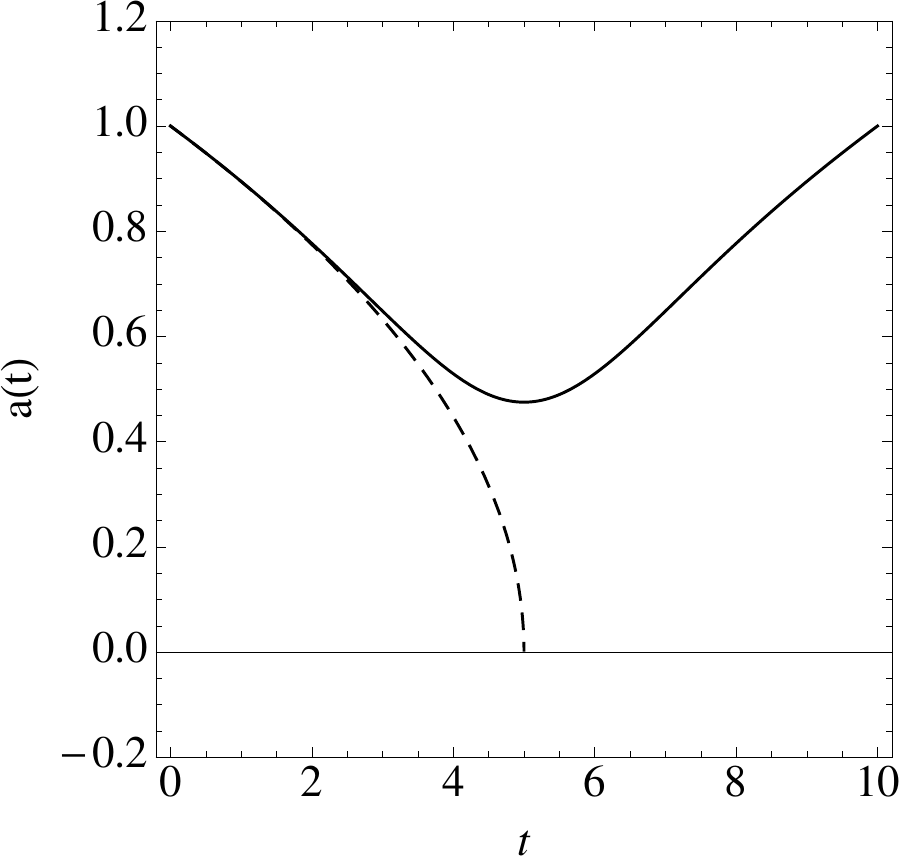}
\hspace{0.5cm}
\includegraphics[type=pdf,ext=.pdf,read=.pdf,width=5.46cm]{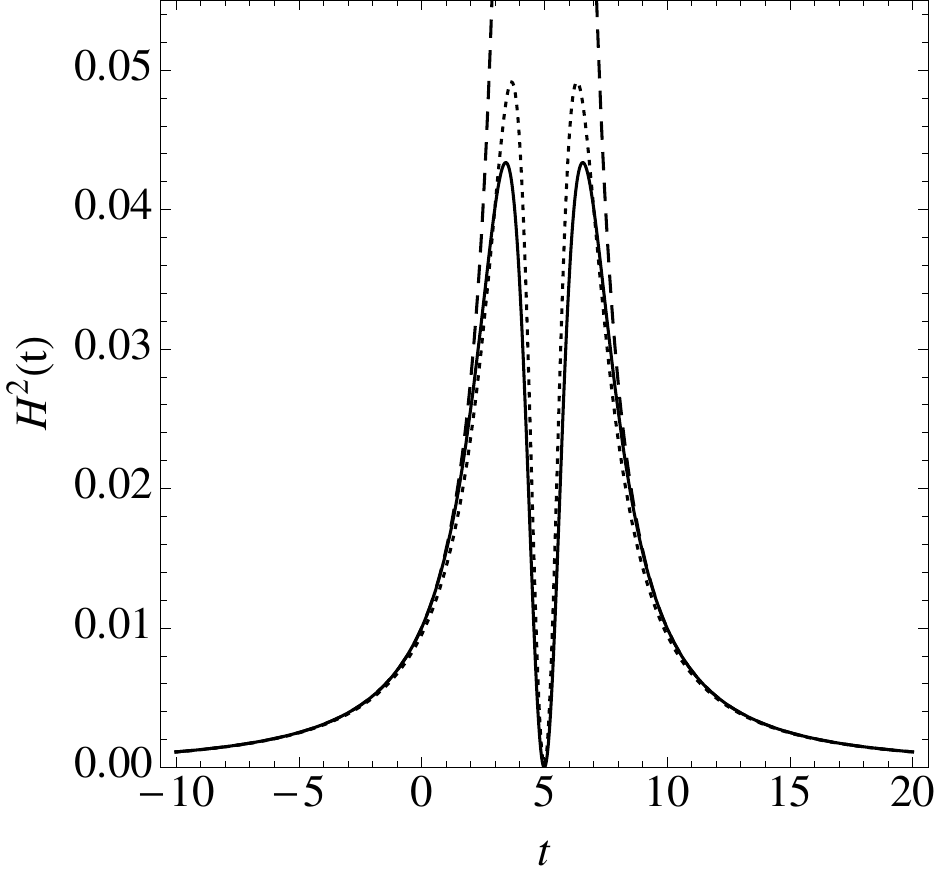}
\hspace{0.5cm}
\includegraphics[type=pdf,ext=.pdf,read=.pdf,width=5.17cm]{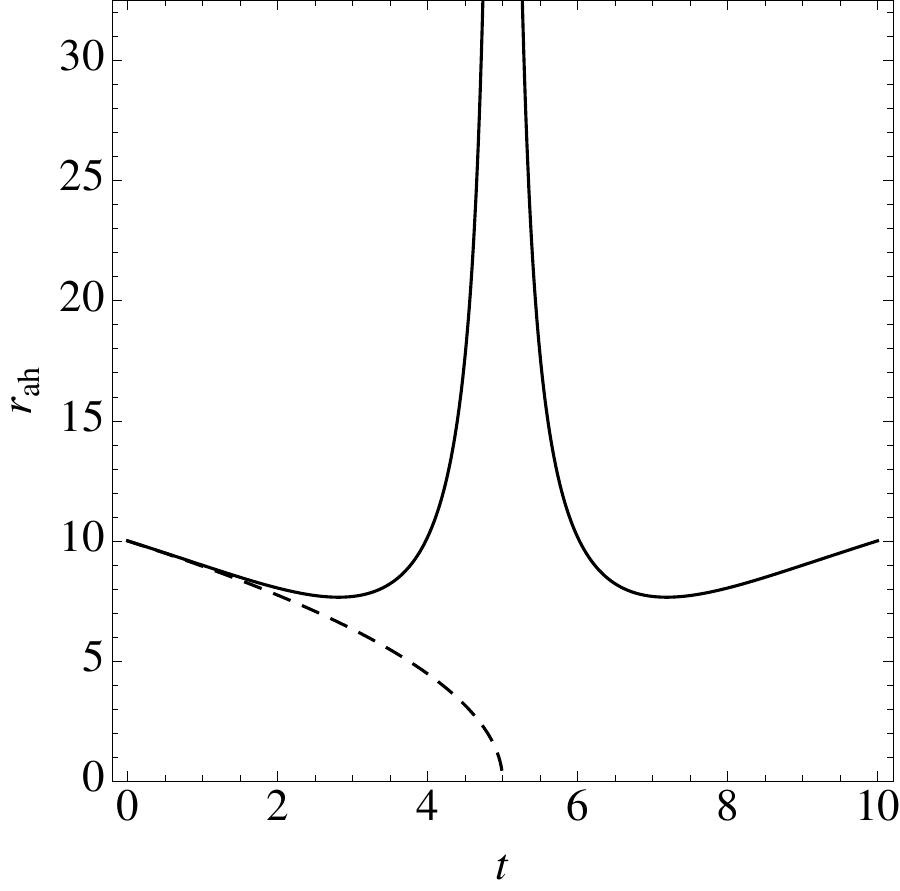}
\caption{Left panel: evolution of the scale factor $a(t)$ for the classical radiation 
collapse (dashed curve) and the semiclassical radiation collapse (solid curve). 
Central panel: as in the left panel for the square of the Hubble rate 
$H^2 = (\dot{a}/a)^2$ and comparison with the effective radiation model with 
$\alpha = 1$ in Eq.~(\ref{H2}) (dotted curve). Right panel: evolution of the radius of the apparent 
horizon $r_{\rm ah}$ for the 
classical radiation collapse (dashed curve) and semiclassical radiation collapse 
(solid curve). Here $t_0 = 5$ and $\Lambda = 1$. See the text for more details.} 
\label{fig1}
\end{figure*}

Let us first study the radiation case where $p=\rho/3$. The classical solution is
\be
a(t)^2 = 
 \left| \frac{t_0 - t}{t_0}  \right| \, ,
\label{arad}
\ee
where $t=t_0$ is the time of occurrence of the singularity. For the theory defined in 
Eq.~(\ref{theory}), it is more convenient to find the solution of the scale factor
with the propagator approach~\cite{broda}, rather than by solving the
counterpart of the Einstein equations~(\ref{einstein-1}). The procedure and the
details of the calculations are reported in Appendix~\ref{s-a-2}.
The final result is
\be\label{a1}
\hspace{-0.3cm}
a^2(t) = \frac{ 2 e^{-\frac{1}{4} \Lambda^2 (t-t_0)^2}}{\Lambda \sqrt{\pi } \,  t_0} +  
\frac{ (t_0 -t) \, \text{erf} \left(\frac{\Lambda (t_0 - t)}{2}\right)}{ t_0}  \, ,
\ee
where ${\rm erf}(z) = 2\int_0^z \exp(-t^2) dt/\sqrt{\pi}$. The classical singularity is 
now replaced by a bounce at $t=t_0$, as can be seen in the left panel of Fig.~\ref{fig1}. 
Following the spirit of Ref.~\cite{bmm}, we can write the effective Einstein equations,
in which $\rho$ and $p$ in Eq.~(\ref{einstein-1}) are replaced, respectively,
by an effective density $\rho_{\rm eff}$ and an effective pressure $p_{\rm eff}$.
The effective pressure is
\be
p_{\rm eff} = - \frac{4}{\kappa^2} \left[ \left( \frac{\dot{a}}{a} \right)^2 
+ 2 \, \frac{\ddot{a}}{a} \right]  \, . \label{pressure}
\ee
$p_{\rm eff}$ is close to the classical value $p = \rho/3$ far from the time $t = t_0$,
while it becomes negative around $t=t_0$.

Asymptotic freedom plays a crucial rule in the kind of approximation we 
are doing. It allows us to use only the two points function (propagator) because 
all the $n$-graviton interactions go to zero at high densities near the bounce.
However, a generic asymptotic freedom is sufficient to remove the singularity, 
but it is not enough to have a bounce. Here, the asymptotic freedom is due to 
a higher derivative form factor, which makes gravity repulsive at very small 
distances. In particular, we would like to stress that the repulsion that causes 
the cloud to expand is not given by the quantum mechanical nature (Heisenberg 
uncertainty) of the collapsing matter in the regime in which gravity vanishes. 
The bounce follows from the dynamics of the system. In terms of the effective 
picture, the bounce comes from the conservation of the (effective) energy-momentum 
tensor: the matter is transformed into a state with $\rho_{\rm eff}+p_{\rm eff} < 0$, 
which is unstable and therefore the bounce is the only available possibility.

The Hubble rate $H=\dot{a}/a$ is shown in the central panel of Fig.~\ref{fig1}. It 
is interesting to compare this $H$ with the one we can obtain from an effective theory
in which we introduce an effective energy density expressed in terms of the radiation 
energy density~\cite{bojo,bmm} 
\be\label{H2}
H^{2} := \frac{\kappa^2}{12} \rho_{\rm eff} 
= \frac{\kappa^2}{12} \rho \left[1- \left( \frac{\rho}{\rho_{\rm cr}}
\right)^{\alpha} \right] \, ,
\ee
where $\rho = \rho_0/a^4$ is the (physical) radiation energy density,
$\rho_0 = 12/(\kappa^2 \, t_0^2)$, and $\alpha$ is a model-dependent parameter
(for instance, $\alpha \approx 1$ in loop quantum cosmology~\cite{bojo}). 
When $\Lambda \sim M_{\rm Pl}$, we
should expect that the critical density $\rho_{\rm cr}$ is of order the
Planck energy density. The plot of $H^2(t)$ for $\alpha = 1$ is shown
in the central panel of Fig.~\ref{fig1}. The modifications induced by quantum effects 
can essentially be incorporated in a new term proportional to $\rho^{2}$, which
is negligible for $\rho\ll\rho_{\rm cr}$ and becomes relevant as $\rho$ approaches 
$\rho_{\rm cr}$. When $\rho=\rho_{\rm cr}$, gravity is turned off, $H = 0$,
and we have the bounce.

The same procedure can be followed to study the collapse of dust. Now the 
classical solution is
\be
\hspace{-0.2cm} 
a(t)^2 =  \left| \frac{t_0 - t}{t_0} \right|^{\frac{4}{3}} \, .
\label{ahd}
\ee
The quantum-gravity-corrected solution is
\be
\hspace{-0.4cm} 
a^2(t) = -
\frac{2  \Gamma \left(-\frac{2}{3}\right) \Gamma
   \left(\frac{4}{3}\right)\!_1F_1\left(-\frac{2}{3};\frac{1}{2};-\frac{(t_0 - t)^2 \Lambda
   ^2}{4} \right)}{ \Lambda^{4/3} \sqrt{3} \pi  t_0^{4/3}}\,  , 
\ee
where $_1F_1$ is the Kummer 
confluent hypergeometric function. Just as in the radiation case, also in the case 
of dust matter there is a bounce at the time $t_0$ in place of classical singularity.

\section{Trapped surfaces and Penrose diagram}

The condition for the formation of trapped surfaces is given by the requirement 
that the surface $R(r,t) = {\rm constant}$ is null; that is, $g^{\mu\nu} (\partial_\mu R)
(\partial_\nu R) = 0$. In our homogeneous marginally bound collapse, this reduces
to $1 - \dot{R}^2 = 0$, and therefore
\be
\label{trapped-surfaces}
r_{\rm ah} = \frac{1}{|\dot{a}|} \, .
\ee
The right panel of Fig.~\ref{fig1} shows the evolution of the radius of the apparent 
horizon $r_{\rm ah}$ for the radiation model, both in the classical and quantum 
scenarios. In the classical case, there is no way to avoid the formation of the 
apparent horizon: the latter forms at the boundary 
of the collapsing cloud at a time $t < t_0$, before the formation of the singularity, 
and then propagates inwards to reach the center at the time of formation of the 
singularity. When the collapsing cloud crosses the Schwarzschild radius, the 
event horizon forms in the exterior spacetime and the formation of a black hole 
as the final stage of the collapse is indicated by the instant of formation of the 
trapped surfaces.

As shown in the right panel of Fig.~\ref{fig1}, the semiclassical scenario is qualitatively 
different. In both the radiation and the dust collapses, the curve $r_{\rm ah}(t)$ is
delayed with respect to the classical model and then reaches a minimum at a time $t_*$, 
where $\ddot{a}=0$ and $\dot{a}$ reaches a maximum value $\dot{a}(t_*)=\dot{a}_*$. 
This leads to the existence of a limiting radius $r_*$
\begin{equation}
r_*=\frac{1}{|\dot{a}_*|} \, .
\end{equation}
If the boundary of the cloud is $r_{\rm b} < r_*$, then no trapped surface forms at any 
time during the collapse. We have thus a threshold mass, below which the 
collapsing matter can always be seen by a distant observer.
Of course, this threshold is related to the scale introduced by the quantum
theory, in our case by $\Lambda$, and therefore can be relevant for objects
of planckian size.

Within this semiclassical scenario, it seems that a black hole, 
strictly speaking, never forms, in the sense that there is no region causally 
disconnected from future null infinity. 
The whole picture can be summarized as follows.
At the beginning, the semiclassical collapse is close to the classical 
scenario. As the matter density increases, the gravitational force becomes weaker. 
In the language of the effective picture, quantum gravity effects become important 
when the physical energy density approaches the critical one and the effective 
energy density goes to zero. In both the semiclassical radiation and dust models, 
we have a bounce, after which the collapse turns into an expansion. Near the 
critical time of the bounce, gravity is weak (it is completely turned off at the time 
$t_0$ of the bounce) and any horizon disappears, at least in the interior solution,
thus leaving the high density region potentially visible to distant observers.
Such a possibility is eventually determined by the form and behavior of the
exterior $r > r_{\rm b}$ spacetime, which we do not know for the full quantum-gravity
theory under consideration. 
However the semiclassical analogy, together with classical models 
matching to generalized Vaidya spacetimes with outgoing radiation, 
and some arguments related to the continuity of the trapping horizon 
suggest that the trapped region disappears also in the exterior.
After the bounce, a new horizon forms, as a consequence of the decrease in 
the matter energy density and the increase of the gravity strength, and then 
disappears for ever at later times, when the radius of the apparent horizon 
exceeds $r_{\rm b}$. The formation/evaporation of the trapped surface is determined 
by the asymptotically free nature of gravity. The second trapped surface forms 
when gravity leaves the asymptotic freedom regime and it becomes strong again.

\begin{figure}
\hspace{0.7cm}
\includegraphics[type=pdf,ext=.pdf,read=.pdf,width=7cm]{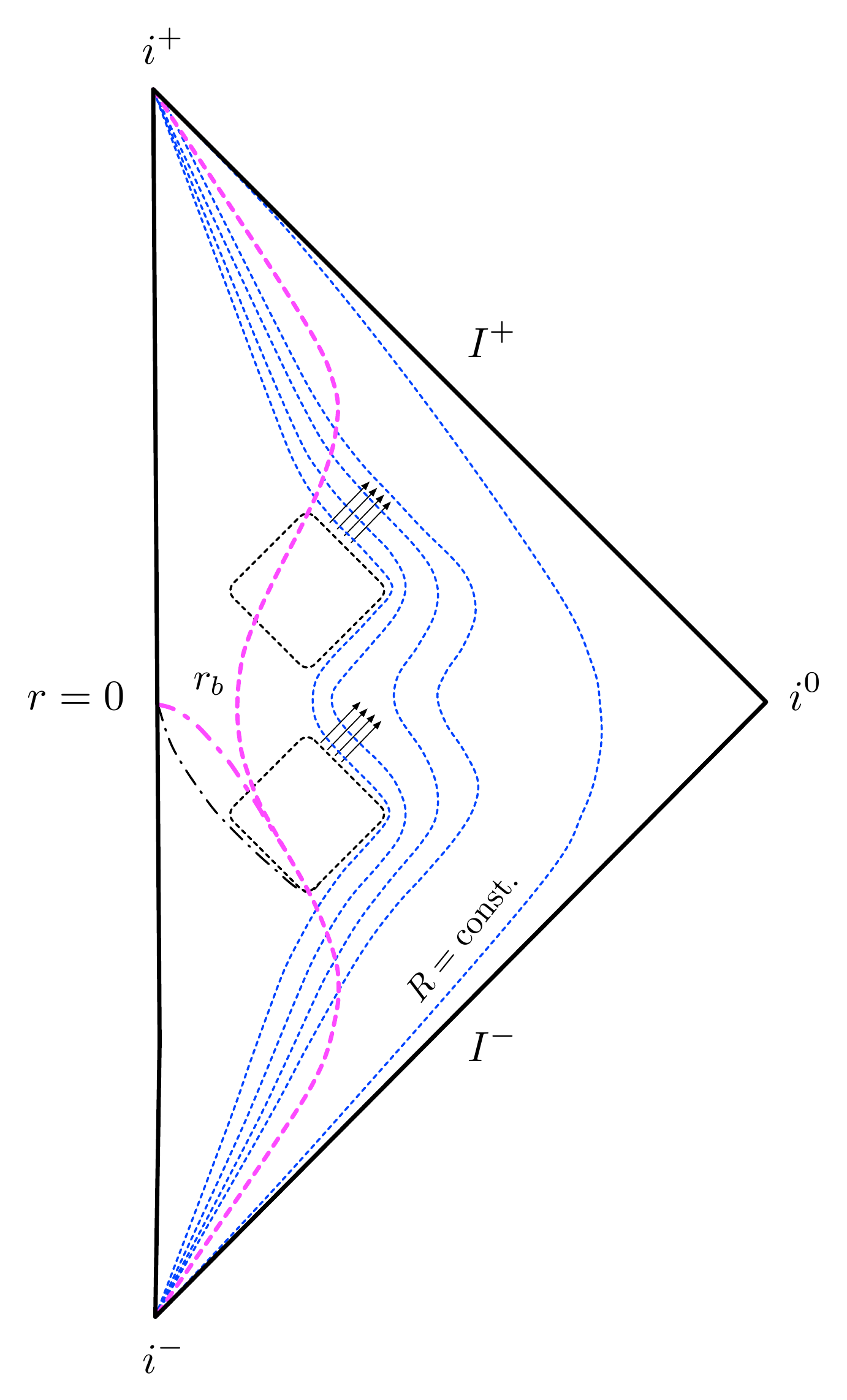}
\caption{Penrose diagram for the semiclassical radiation collapse model
described by Eq.~(\ref{a1}). The magenta thick-dashed line is the curve 
of the radius of the boundary of the collapsing object (the timeline curve 
defined by $R_{\rm b}(t) = r_{\rm b} a(t)$), 
while the black thin-dashed lines represent the trapped surface. 
In the classical model, there is no bounce; at a certain time, the behavior 
departs from the semiclassical solution, as shown by the dotted-dashed 
lines (magenta thick line for $r_{\rm b}$, black thin line for $r_{\rm ah}$). 
The blue-dotted lines are curves of constant radial coordinate.}
\label{Penrose} 
\end{figure}

The Penrose diagram is shown in Fig.~\ref{Penrose}. Unlike the classical 
case where $t_0$ represents the singularity time, the conformal diagram for 
the collapse model extends to $t$ going to future infinity. In the semiclassical 
picture, the trapped region develops as in the classical regime, but then 
disappears at the bounce due to the semiclassical corrections, and is 
accompanied by a second trapped region in the corresponding expanding 
phase. Our theory has small departures from classical general relativity at 
low density/curvature and no superluminal motion. The key-point to
understand the Penrose diagram, and in particular the destruction of the 
horizon, is the following. Interior homogeneous solutions matched with
an exterior vacuum Schwarzschild spacetime hold in the special case of
classical general relativity with a cloud of dust. Beyond general relativity,
the Birkhoff's theorem does not usually hold. 
However, we can recast the quantum-gravity theory in a semiclassical
effective theory that describes a fluid that in general is not dust and violates
the energy conditions.
In the general case, therefore including in general relativity but without dust, 
the matching has to be done with a generalized Vaidya solution, which represents 
a spacetime with ingoing or outgoing null flux of energy. 
While here we have only the interior $r < r_{\rm b}$ metric, the
external part in the effective picture
is surely a Vaidya metric. In other words, our collapsing
object presumably has a lot of ``hairs''. If we want to see the collapse in terms
of the effective picture, in which the Einstein equations are satisfied by 
an effective fluid covering the whole spacetime, the exterior solution
should be an ingoing flux of negative energy. It is this external ingoing 
flux that allows for the destruction of the horizon. As the theory has small
deviations from classical general relativity at low energy density/curvature,
it is clear that this flux is very low and therefore that the lifetime of the
trapped surface -- as measured by a distant observer -- must be long for
an astrophysical object with $M/\Lambda \gg 1$ (see next section for more 
details).

The above picture for gravitational collapse seems plausible,
because the matching with the Vaidya exterior is a well known procedure and
the interpretation appears quite natural. However, without knowing the exact
form of the exterior solution, one can not in principle exclude other scenarios.
For example, the outward flux of energy may
be irrelevant, thus leaving an exterior that is almost vacuum and
therefore the gravitational collapse may produce an ordinary black
hole with an almost constant mass $M_{\rm in}$ given by the gravitational mass
of the collapsing cloud. After the bounce, the expanding cloud 
would then be confined inside the Schwarzschild radius and 
would evolve into a new (expanding) universe. 
In this case, the Penrose diagram could look like one of those reported 
in Ref.~\cite{boulware}. 

Universes created as offspring of collapse to black holes have been 
considered in the literature
\cite{baby}.
This kind of scenario can generally be obtained analytically by means 
of a cut and paste procedure in which a singular manifold, such as the
Schwarzschild black hole, can be extended beyond the singularity by removing 
the same and sewing the spacetime to a new non singular manifold 
describing an expanding baby universe.
However, as far as we are aware, even this procedure is feasible only 
in very simple examples and becomes highly non trivial if one wishes
to consider the dynamical setup.
Typically the matching involves continuity of the first and second 
fundamental forms across some hypersurface. To have the chance 
to fulfill these requirements one needs a large enough number of 
free parameters, which is not the case of marginally bound collapse.

Therefore, despite the general appeal that such a solution may have, 
one is faced with a lot of technical difficulties plus the important objection 
that the fact that a certain manifold can be constructed by hand does not imply 
necessarily that it is realistic. 
In this sense, an interior solution in the form of a scale factor $a$ obtained 
from a well posed theory and valid globally without any junction 
appears to be more natural choice. 
Indeed a selection principle as the one proposed by Smolin~\cite{smolin}
is practically very difficult to achieve even if one neglects 
microphysics and considers only some effective theory 
of gravity as we do here.

\section{Astrophysical objects}

A model consistent with observations must be able to explain the super-massive 
black hole candidates in galactic nuclei, and therefore the trapped surface 
formed in the collapse for an object with mass $M \sim 10^5-10^9$~$M_\odot$ must 
survive for a time at least comparable to the age of the Universe. 
For a heavy ($M/\Lambda \gg 1$) astrophysical object, it is natural to expect that
this is indeed the case, because the theory has small deviations from classical
general relativity at low curvature and therefore the ingoing flux responsible for the
destruction of the horizon must be very low; that is, the lifetime of the trapped 
surface must be long. For a comoving observer, the timescale of the collapse is 
of order the dynamical timescale $\kappa M \sim 1 \; (M/10^6 \; M_\odot)$~s. 
For a distant observer, the timescale is longer, as a consequence of the gravitational 
redshift. The exact calculation of the lifetime of the trapped surface
would require the knowledge of the metric in the whole spacetime, while
in our case we have only the interior solution. An estimate of this time
interval can be obtained from the velocity of shrinkage of the horizon~\cite{frolov}
\be
\sigma = \left(\frac{dr}{dv}\right)_{\!\!g_{00} = 0} \hspace{-0.0cm} ,
\ee
where $v=t+r$ is the advanced time 
and $g_{00}$ is the temporal component of 
the metric, while $g_{00} = 0$ defines the apparent horizon. 
The shrinkage vanishes for the classical black hole case and the lifetime of the
horizon is thus infinite in this case. As discussed in the previous section, one
can also see the evaporation of the trapped surface in terms of an effective 
picture, in which the horizon is destroyed by an ingoing flux of negative
energy. As we can play only with two mass scales, $\Lambda$
and $M$, we may guess that the order of magnitude is given by $\Lambda/M$,
or that it is given by an expansion in $\Lambda/M$ and therefore even more 
suppressed. The lifetime of the trapped surface with respect to the distant observer 
is, in the more conservative case with $dr/dv \sim \Lambda/M$,
\be
\tau \sim \frac{\kappa M}{dr/dv} \sim \kappa M \left(\frac{M}{\Lambda}\right) \, ,
\ee
which is anyway much longer than the age of the Universe (for the Sun, 
$M/\Lambda \sim 10^{38}$ if $\Lambda$ is of order the Planck mass).

Our theoretical model (homogeneous cloud of dust or radiation) is very simple,
but it is easy to figure out how the picture might change in a more realistic scenario,
at least qualitatively. It is natural to expect that the cloud is inhomogeneous, with
a density profile monotonically decreasing in the outwards radial direction. 
Our results for the homogeneous case should hold along the central shell 
of the collapsing cloud, while at larger radii, since the density should be lower, 
one expects a smaller deviation from the standard general relativistic case. 
The bounce will remain, but at larger radii gravity may still be strong 
and therefore the instant of the bounce may not be visible to the distant 
observers (in the language of the Penrose diagram 
in Fig.~\ref{Penrose}, the two disconnected gray areas representing the trapped 
regions would be a single region).

An inevitable effect is an apparent mass loss of astrophysical black hole 
candidates. This is a consequence of asymptotic freedom and, strictly speaking, 
does not require the presence of the bounce. In other words, the only necessary ingredient
is that gravity becomes very weak at high densities, while a negative effective pressure
is not strictly necessary. As it is more clear in terms of the effective Einstein equations, 
the effective density decreases when the physical density approaches the critical 
one and increases after the bounce. The gravitational mass seen by a distant 
observer, i.e. $F_{\rm eff}(r_{\rm b}, t)/2$ where $r_{\rm b}$ is the boundary, does the 
same (at the bounce $\dot{a} = 0$, so $\rho_{\rm eff} = 0$). A collapsing object
should thus appear as a black hole candidate with a time varying mass. Such a 
prediction is robust, even if the exact behavior may depend on both the theoretical 
framework (e.g. the choice of the form factor) and the astrophysical content 
(matter equation of state, initial conditions, etc.), and should be seen as an 
apparent mass loss of black hole candidates. 
If the apparent mass loss rate were to be relatively low and diluted for a 
long time, there might be the chance to observe it as an increase in the 
orbital period of a black hole binary. The future discovery of black hole binaries 
with a pulsar companion can presumably put the strongest constraints on such 
a possibility. On the other hand, if the apparent mass loss rate were high and for 
a relatively short time, resulting in a sudden disappearance of the object,
the phenomenon may easily generate hypervelocity stars, i.e. stars which are 
observed with velocities of order 1,000~km/s but whose origin is not yet 
clear~\cite{h-star}. Indeed, if the stellar companion were in a close high velocity orbit
and could not feel the gravitational force of the black hole candidate for a while, 
it would escape with a velocity equivalent to its orbital velocity. The phenomenon 
may be particularly interesting to produce hypervelocity neutron stars.

\section{Summary and Conclusions}

In this paper, we have 
studied the homogeneous collapse of a cloud in a super-renormalizable and 
asymptotically free theory of gravity. The final singularity of classical general 
relativity is removed and replaced by a bounce. 
Unitarity is necessary to have a ``good'' theory, but it is irrelevant for the presence
of the bounce, as it can be seen in conformal gravity (which is asymptotically free, 
is not unitary, and predicts the bounce \cite{frolov}). A generic asymptotic freedom
is sufficient to remove the singularity, but it is not enough to have a bounce
(see the case of QCD, where the asymptotic freedom is given by the matter 
content). The bounce requires a repulsive gravitational force at high densities.
The key-point is therefore the form factor, which is related to the propagator 
and to the effective potential of the theory. Asymptotic freedom due to a 
higher derivative form factor introduces an effective negative pressure, which
is responsible for the bounce. As in several quantum gravity approaches 
corrections to classical general relativity make gravity repulsive at very high 
densities, independently of the exact UV completion, the prediction of the
bounce is much more general and presumably holds in a larger class of theories.
We indeed note that a bounce replacing the classical singularity in the
gravitational collapse was previously found in different contexts~\cite{casadio}.

We argue that in these theories black holes, strictly speaking, never form.
The Penrose diagram is shown in Fig.~\ref{Penrose}. The theory has small 
deviations from classical general relativity at low densities/curvature and no
superluminal motion. The shrinkage of the external horizon is possible
due to the matching of the effective solution describing the quantum corrections 
with a Vaidya sapcetime in the exterior manifold. For massive astrophysical
objects, deviations from general relativity are tiny and we can therefore
expect that the lifetime of the trapped surfaces is long for an observer at infinity.
Astrophysical black hole candidates may thus be objects with a temporary trapped
surface, but they would be interpreted as black holes if the observational
time scale is much shorter than the lifetime of the horizon.
However, as we have derived only the interior solution in the full quantum-gravity theory, 
we cannot really conclude that this is only possible scenario. 
The effect of the exterior Vaidya solution may very well be negligible 
in the dynamics of the collapse, even when integrated for a very long time. 
If this were to be the case, the collapse would produce a black hole with
the usual Schwarzschild event horizon and the matter cloud, re-expanding 
after the bounce, would evolve into a new universe.

While our work neglects Hawking radiation, it however suggests a simple way to
resolve the information paradox. In the first scenario, the information is trapped 
inside the apparent horizon and released when the latter eventually evaporates. 
It is an example of ``complete evaporation scenario'' according to the terminology 
of Ref.~\cite{infopara}. In the second case with a new universe, the spacetime
decompose into two regions and the information is stored in the new universe.
This is the ``baby universe solution''~\cite{infopara}.


\begin{acknowledgments}
We thank Antonino Marciano for reading a preliminary version of this
manuscript and providing useful feedback.
This work was supported by the NSFC grant No.~11305038, the Innovation 
Program of Shanghai Municipal Education Commission grant No.~14ZZ001, 
the Thousand Young Talents Program, and Fudan University.
\end{acknowledgments}

\vspace{-0.1cm}


\appendix

\section{Ultraviolet and Infrared properties of the theory \label{s-a-1}}

We consider a particular representative theory of the following generic class
\be
\hspace{-0.3cm}
\mathcal{L} = 2 \kappa^{-2} \sqrt{|g|} 
\Big[R - G_{\mu\nu} \frac{ V(-\Box/\Lambda^2)^{-1} -1}{\Box} 
R^{\mu\nu} \Big] \, ,
\label{theory-app}
\ee
where $G_{\mu\nu}$ is the Einstein tensor and $V(-\Box/\Lambda^2)$ 
is an entire function.
At classical level, all the corrections to the Einstein-Hilbert action are suppressed by 
$1/\Lambda$. Since $\Lambda$ is expected to be of order the Planck mass, at low
energies the theory reduces to the Einstein one. At quantum level,
the introduction of non-local operators in the action could leads to strong
non-localities generated by the renormalization group flow towards the infrared,
in disagreement with observations. This is not the case for the Lagrangian in 
Eq.~(\ref{theory-app}), as we can see from the Donoghue argument~\cite{dono}.

The Donoghue result can be summarized as follows. If we start from a general 
covariant theory of gravity involving a Taylor expandable classical action,
at quantum level we find analytical as well as non-analytical finite universal 
contributions coming from one loop diagrams. If only massless particles are 
propagating, around the flat spacetime the non-analytical contribution has the 
form
\be
\kappa^2 k^2 \log(-k^2) \, .
\label{A2}
\ee
If we couple the theory to massive particles, we also have
\be
\kappa^2 k^2 \sqrt{\frac{m^2}{-k^2} } \, .
\label{A3}
\ee
The analytical contributions are instead integer powers of the momentum $k$.
This result also applies to our theory because the action involves only 
{\em entire functions}. The logarithmic non-analytic contribution to the one loop 
effective action is related to the quantum corrections due to long distance 
effects of massless particles. In QED, one can see something similar in the 
photon vacuum polarization,
\be
k^2 \Pi(k^2) \sim k^2 \left(  \frac{1}{6\epsilon} - \frac{ \gamma_E}{6} + f(k) \right) \, .
\ee
The other finite contributions to the one loop amplitude are a series of analytic 
and sub-leading operators in the infrared regime $k \rightarrow 0$,
\be
\lim_{k \rightarrow 0} \sum_n R^n = 0\, , 
\ee
where $R$ is a general local curvature invariant. 
On the other hand,
the non-analytic contribution (\ref{A2}) is divergent in the IR and gives corrections to 
the gravitational potential~\cite{dono}. We find the same situation for the photon 
vacuum polarization in QED if we take the limit of zero electron mass. The 
finite non-analytic contributions coming from the divergent integrals in a 
massless theory and in our theory read 
\be
&&\int  \frac{d^{4-\epsilon} k}{p^2(p - k)^2} 
 = \Gamma(\epsilon/2) \, ( - k^2)^{\epsilon/2} =   
 \label{finiteDivIR}
  \\
&& \hspace{0.0cm}= \left(  \frac{2}{\epsilon} - \gamma_E + O(\epsilon) \right) 
 \left( 1 - \frac{\epsilon}{2} \ln ( - k^2/\mu^2)+ O(\epsilon^2)  \right) =\nonumber\\
&& \hspace{0.0cm} = \frac{2}{\epsilon} - \ln \frac{- k^2}{\mu^2} + O(\epsilon) \nonumber \\
&& \hspace{0.0cm} = \log \frac{\Lambda_{\rm UV}^2 }{\mu^2} - \ln \frac{- k^2}{\mu^2}+ O(\epsilon) 
 \equiv - \ln \left( \frac{- k^2}{\Lambda_{\rm UV}^2 }\right) + O(\epsilon) \nonumber
 \, .
\ee
Finite contributions are analytical operators $O(1/l^4)$ in $D=4$ (for example 
$R^2 \sim 1/l^4$) and then they do not affect the infrared theory even in our 
theory because only {\em entire functions} are present in the action. 
In general, the relevant 
quantities to get the one-loop effective action are~\cite{GBV}
\be
{\rm Tr} \ln \hat{\Box} \, , \qquad
 \nabla_{\mu_1} \dots \nabla_{\mu_p}  \frac{\!\!\!\hat{1}}{\Box^n} 
 \, \delta(x,y) \Big|_{y = x} \, .
\label{quantity}
\ee
In the coincidence limit, the logarithmic divergent contributions of the universal 
quantities (\ref{quantity}) have the structure in~(\ref{finiteDivIR}). Other possible finite 
contributions are instead analytical and polynomial in the momentum.

In our higher derivative theory, the ultraviolet behavior is different with respect to the 
Einstein theory and it depends on the details of the effective action. The 
renormalization group has a non-linear behavior going from the particular 
ultraviolet regime associated with our regularized theory to the universal Einstein 
regime in the infrared. 
The Donoghue
result shows that the infrared modifications are independent of the nature of 
the fundamental higher derivative theory and then equivalent to those of the Einstein theory
(\ref{A2}) and  (\ref{A3}).

So far we have been generic. However, 
the theory is uniquely specified once the form factor is fixed, because it does not 
receive any renormalization at quantum level. In other words, the ultraviolet 
theory is dominated by the bare action. For simplicity, let us assume the following 
Tomboulis form factor
\be
V^{-1}(z) = e^{H(z)} = \left| z^{\gamma  + 1} \right|  \, 
e^{\frac{1}{2} \left[ \Gamma \left(0, z^{2 \gamma  + 2} \right)+\gamma_E  \right] } \, .
\ee
The theory is therefore completely specified and the asymptotic behavior in the ultraviolet 
regime reads
\be
\mathcal{L}_{\rm UV}  \approx \frac{2 \kappa^{-2} 
e^{\gamma_{\rm E}/2} }{\Lambda^{2 \gamma +2}} 
\left( \frac{1}{2} R\, \Box^{\gamma } R - R_{\mu \nu} 
\Box^{\gamma } R^{\mu \nu}  \right) \,  ,
\ee
which depends only on the integer exponent $\gamma$. In this paper, we used the form 
factor suggested by string field theory to make easy the classical analysis. 
However, the main result of the paper is insensitive to the details of the theory.

The form factor $V(z)$ must have no extra poles 
in the whole complex plane, but it is also constrained to have a renormalizable 
or finite theory in the ultraviolet regime. This leaves us with a class of theories 
each of them super-renormalizable because of the following  reasons:
\begin{enumerate}
\item only a finite number of couplings is renormalized, 
\item only a finite number of diagrams is divergent. 
\end{enumerate}
On the other hand, at the phenomenological level the form factor could be 
experimentally constrained, for example measuring the corrections to the gravitational 
potential, or hypothetically measuring a cross section in a scattering process at 
high energy. Since $V(z)$ is an entire function, there are no ghosts and no 
tachyons, independently of the number of time derivatives present in the action. 
This is the main reason to introduce a non-polynomial Lagrangian.

Concerning the Lorentz invariant scale $\Lambda$, there is no fine tuning and 
it is not subject to infinite or finite (non analytic) renormalizations. It is only 
constrained to be a large mass scale by astrophysical or cosmological 
observations and it is natural to expect $\Lambda$ of order 
the Planck mass. As a consequence of (\ref{quantity}) and (\ref{finiteDivIR}),
in the 
ultraviolet regime the one loop corrections to the classical theory read 
\be
R \ln (- \Box) R \,\,\,\,  \mbox{and} \,\,\,\,  R_{\mu \nu} \ln (- \Box) R^{\mu \nu}.
\ee
There are no finite renormalizations of $\Lambda$ because all the finite and 
infinite corrections to the operators  $R \, \Box^n R$ and $R_{\mu \nu} \, \Box^n R^{\mu \nu}$
are absorbed in the running couplings of the theory and not in the scale 
$\Lambda$. However, this operators do not move the poles in the propagator, 
because they are suppressed by the form factor $V(z)$.

\section{Solution of the scale factor via the propagator approach \label{s-a-2}}


The energy-momentum 
tensor in comoving coordinates for the generic spherically symmetric
metric describing collapse given in Eq.~(\ref{eq1}) is given by 
\begin{equation}
T^\mu_\nu={\rm diag}\{\rho(r,t), p_r(r,t), p_\theta(r,t), p_\theta(r,t) \} \; .
\end{equation} 
Einstein's equations relate the metric functions to the matter content and are given by
\begin{eqnarray}\label{rho}
  p_r &=&-\frac{\dot{F}}{R^2\dot{R}} \; , \qquad
  \rho=\frac{F'}{R^2R'} \; , \\ \label{nu}
  \nu'&=&2\,\frac{p_\theta-p_r}{\rho+p_r}\frac{R'}{R}
  -\frac{p_r'}{\rho+p_r} \; ,\\ \label{Gdot}
  \dot{G}&=&2\,\frac{\nu'}{R'}\dot{R}\,G \; ,
\end{eqnarray}
where the $'$ denotes a derivative with respect to $r$, and the $\dot{}$ denotes a 
derivative with respect to $t$. The function $F(r,t)$ is called Misner-Sharp mass, 
and in general is
\begin{equation}\label{misner}
F=R(1-G+e^{-2\nu}\dot{R}^2) \; .
\end{equation}
In the homogeneous marginally bound case, from the first of Eq.~(\ref{rho}) it 
follows that $F$ is a function of $r$ only and the matching to the exterior vacuum 
Schwarzschild spacetime is always possible. Furthermore Eq.~(\ref{nu}) reduces 
to $\nu'=0$ and we can always choose the time coordinate in such a way that
$\nu = 0$. Integration of Eq.~(\ref{Gdot}) is then trivial and gives $G = 1+f(r)$ 
and in the marginally bound collapse case we shall take the free integration 
function $f$ to be zero. The system is then fully specified once a gauge is fixed 
for the scale. This is usually done by fixing the scale at the initial time. 
It is common to define $R(r,t)=r a(t)$ with $a(0)=1$, so that to solve the
system we only need to find the scale factor $a(t)$ by solving the corresponding
field equations. Here, we use instead the propagator approach of Ref.~\cite{broda}.

We first write the metric as a flat Minkowski background plus a fluctuation 
$h_{\mu \nu}$,
\be 
g_{\mu \nu} &=& \eta_{\mu \nu} + \kappa \, h_{\mu \nu} \, , \nonumber\\
ds^2 &=& dt^2 - a(t)^2 dx^i dx^j \delta_{i j} \, ,  
\ee
where $\eta_{\mu \nu} = {\rm diag}(1,-1,-1,-1)$. The conformal scale factor $a(t)$ 
and the fluctuation $h_{\mu\nu}(t, \vec{x})$ are related by the following 
relations~\cite{broda}:
\be
&& \hspace{-0.5cm}
a^2(t) = 1 - \kappa h(t) \,  , \,\,
h(t=t_0) = 0 \, , \,\, 
 g_{\mu \nu}(t=t_0) = \eta_{\mu\nu} \, , \nonumber\\
&& \hspace{-0.5cm}
h_{\mu \nu}(t, \vec{x} ) = h(t) \, {\rm diag}(0, \delta_{i j} ) 
\equiv h(t) \, \mathcal{I}_{\mu \nu} \, .
\label{ah}
\ee
After a gauge transformation, we can rewrite the fluctuation in the usual harmonic 
gauge
\be
&& h_{\mu \nu}(x) \rightarrow h^{\prime}_{\mu \nu}(x) = 
h_{\mu \nu}(x)+ \partial_{\mu} \xi_{\nu} + \partial_{\nu} \xi_{\mu} \, , 
\nonumber\\
&& \xi_{\mu}(t) = - \frac{3 \kappa}{2} \, 
{\rm diag}\left( \int_0^{t} h(t') dt',0,0,0  \right) \, . 
\ee
The fluctuation now reads 
\be
&& h^{\prime}_{\mu \nu}(t, \vec{x} ) = 
h(t) \, {\rm diag} ( - 3, \delta_{i j} ) \, , \nonumber\\
&& h^{\prime \, \mu}_{\mu}(t, \vec{x} ) = - 6 h(t) .
\ee
We can then switch to the standard gravitational ``barred" field 
$\bar{h}^{\prime}_{\mu \nu}$ defined by 
\be
\bar{h}^{\prime}_{\mu \nu} = {h}^{\prime}_{\mu \nu} - \frac{1}{2} \eta_{\mu \nu} \, 
h^{\prime \, \lambda}_{\lambda} 
=  - 2 h(t) \, \mathcal{I}_{\mu \nu} \, ,
\ee
satisfying $\partial^{\mu} \bar{h}^{\prime}_{\mu \nu} = 0$.
The Fourier transform of $\bar{h}^{\prime}_{\mu \nu}$ is
\be
\tilde{\bar{h}}^{\prime}_{\mu \nu}(E, \vec{p}) 
= -2 \tilde{h}(E) (2 \pi)^3 \delta^3(\vec{p}) \, \mathcal{I}_{\mu \nu} \, . 
\label{FTh}
\ee

The classical solution for the cosmological 
scenario in the radiation fluid model is
\be
a(t)^2 = 
 \left| t/t_0  \right| \, ,
\label{arad}
\ee
where $t=0$ is the singularity time. With the solution~(\ref{arad}), we can 
compute the Fourier transform $\tilde{h}(E)$ defined in~(\ref{FTh}) 
\be\label{eq-htilde}
\tilde{h}(E) = \frac{2}{\kappa \, t_0 E^2} + \frac{2 \pi }{\kappa}  \delta(E) \, .
\ee
We can obtain the same solution~(\ref{eq-htilde}) from the classical propagator 
if we properly introduce a dimensionless fictitious source in the momentum 
space. We can then extend this procedure to the theory defined in 
Eq.~(\ref{theory-app})~\cite{broda}.
The gauge independent part of the graviton propagator for the theory~(\ref{theory-app})
and energy tensor $\tilde{T}^{\rho\sigma}(p)$ is (see e.g. the third paper in~\cite{sn})
\be 
\hspace{-1cm}
&& \mathcal{O}^{-1}_{\mu \nu\rho\sigma}(p) = 
\frac{V(p^2/\Lambda^2) }{ 
p^2}   
\left( P^{(2)}_{\mu \nu\rho\sigma} - 
\frac{1}{2} P^{(0)}_{\mu \nu\rho\sigma}  \right) 
\nonumber\\
\hspace{-1cm}
&& \Longrightarrow \, 
\bar{h}^{\prime}_{\mu \nu}(x) = 
\kappa \! \int \! \frac{d^4 p}{(2 \pi)^4}
\mathcal{O}^{-1}_{\mu \nu\rho\sigma}(p) 
\tilde{T}^{\rho\sigma}(p) \, e^{i p x} \, ,
\label{propgauge2} 
\ee
where $P^{(2)}_{\mu \nu\rho\sigma}$ and 
$P^{(0)}_{\mu \nu\rho\sigma}$ are the graviton projectors,
\be
&& P^{(2)}_{\mu\nu\rho\sigma} =
\frac{1}{2} \left(\theta_{\mu\rho}\theta_{\nu\sigma} 
+ \theta_{\mu\sigma}\theta_{\nu\rho}\right) 
- \frac{1}{3} \theta_{\mu\nu}\theta_{\rho\sigma} \, , \nonumber\\
&& P^{(0)}_{\mu \nu\rho\sigma} =
\frac{1}{3} \theta_{\mu\nu}\theta_{\rho\sigma} \, , 
\ee 
and
$\theta_{\mu\nu} = \eta_{\mu\nu} - k_\mu k_\nu/k^2$.
Therefore
\be
&& h(t) = \kappa
\int \frac{d^4 p}{(2 \pi)^4} 
\frac{1}{p^2 V^{-1}(p^2/\Lambda^2)} \, \tilde{\rho}(E, \vec{p} ) \,
e^{ i p x} = \nonumber\\
&& \hspace{0.7cm}
=  \kappa \int \frac{d E}{2 \pi} \frac{1}{E^2 V^{-1}(E^2/\Lambda^2)} 
\, \tilde{\rho}(E) e^{i E t} \, . 
\label{htg}
\ee
For $V(p^2/\Lambda^2) =1$, we recover the classical case and the solution for 
$h(t)$ is exactly~(\ref{arad}) if we use the distribution
\be
\hspace{-0.6cm}
\tilde{\rho}(E, \vec{p} ) = 
 \underbrace{ \left( \frac{2}{\kappa^2 t_0}  
+ \frac{2 \pi}{\kappa^2} E^2 \delta(E) \right) }_{\tilde{\rho}(E)}
(2 \pi)^3 \delta^3(\vec{p}) \, . 
\label{Frho}
\ee
We can then use this distribution for the form factor $V(z)^{-1} = e^z$. The fluctuation 
resulting from the integral~(\ref{htg}) now becomes 
\be
\hspace{-0.7cm}
\kappa \, h(t) = 1 - a^2(t) = 1 -   \frac{ 2 e^{-\frac{1}{4} 
\Lambda ^2 t^2}}{\Lambda \sqrt{\pi } \,  t_0}  -  
\frac{ t \, \text{erf} \left(\frac{\Lambda  t}{2}\right)}{ t_0}
   ,
\label{hsft}
\ee
where ${\rm erf}(z) = 2\int_0^z \exp(-t^2) dt/\sqrt{\pi}$. 
The solution is not a gauge artifact because we use the gauge independent
projected propagator.
Since we are interested in gravitational collapse rather than the cosmological 
solution, we replace the time coordinate $t$ with $-t+t_0$ to have the classical 
singularity at $t = t_0$ and the initial time at $t=0$. So
\be
\hspace{-0.5cm}
a^2(t) = \frac{ 2 e^{-\frac{1}{4} \Lambda^2 (t-t_0)^2}}{\Lambda \sqrt{\pi } \,  t_0} +  
\frac{ (-t+t_0) \, \text{erf} \left(\frac{\Lambda (- t+t_0)}{2}\right)}{ t_0}  \, .
\ee

The same procedure can be followed to study the collapse of pressureless matter, i.e. dust.
In the dust case, the classical solution for the cosmological scenario is given by
\be
\hspace{-0.2cm} 
a(t)^2 =  \left| t/t_0 \right|^{\frac{4}{3}} \, ,
\label{ahd}
\ee
and the correct fictitious distribution to get the perturbative solution starting 
from the propagator reads  
\be
\hspace{-0.4cm} 
\tilde{\rho}(E, \vec{p} ) = 
\left( \!
\frac{4 \,  \Gamma \left(\frac{4}{3}\right)}{\sqrt{3} t_0^{4/3} | E |^{\frac{1}{3}} }+2 \pi 
   E^2 \delta ( E ) \right) \! 
   \frac{(2 \pi)^3}{\kappa^2} \delta^3(\vec{p}) \, ,
   \nonumber\\
\label{rhoD}
\ee
where $\Gamma$ is the Euler gamma function.
Using the distribution in~(\ref{rhoD}), we can find the modified solution for 
the gravitational fluctuation~(\ref{htg}). From $a^2(t)=1-\kappa h(t)$, we get
\be
&& \hspace{-1cm}
a^2(t) = -
\frac{2  \Gamma \left(-\frac{2}{3}\right) \Gamma
   \left(\frac{4}{3}\right) \, _1F_1\left(-\frac{2}{3};\frac{1}{2};-\frac{1}{4} t^2 \Lambda
   ^2\right)}{ \Lambda^{4/3} \sqrt{3} \pi  t_0^{4/3}}\,  , 
\ee
where 
$_1F_1$ is the Kummer confluent hypergeometric function. The solution for 
the gravitational collapse scenario is obtained by replacing $t$ with $-t+t_0$,
as done for the radiation model.

\begin{figure*}
\hspace{-0.3cm}
\includegraphics[type=pdf,ext=.pdf,read=.pdf,width=5.43cm]{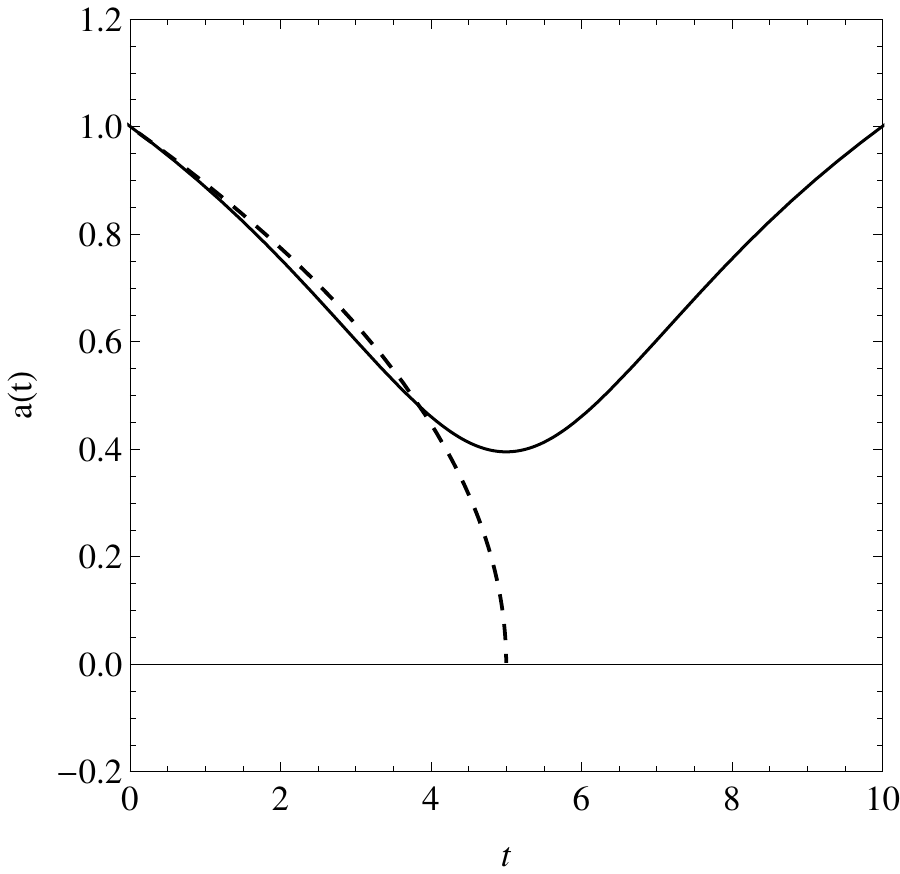}
\hspace{0.5cm}
\includegraphics[type=pdf,ext=.pdf,read=.pdf,width=5.46cm]{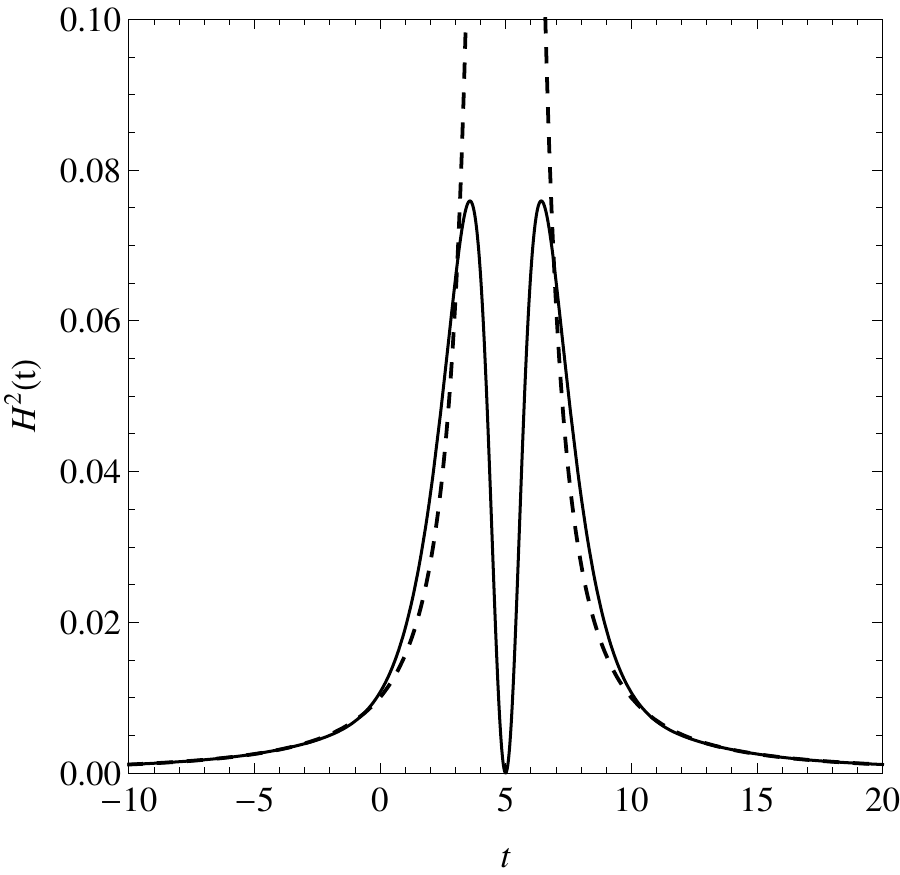}
\hspace{0.5cm}
\includegraphics[type=pdf,ext=.pdf,read=.pdf,width=5.17cm]{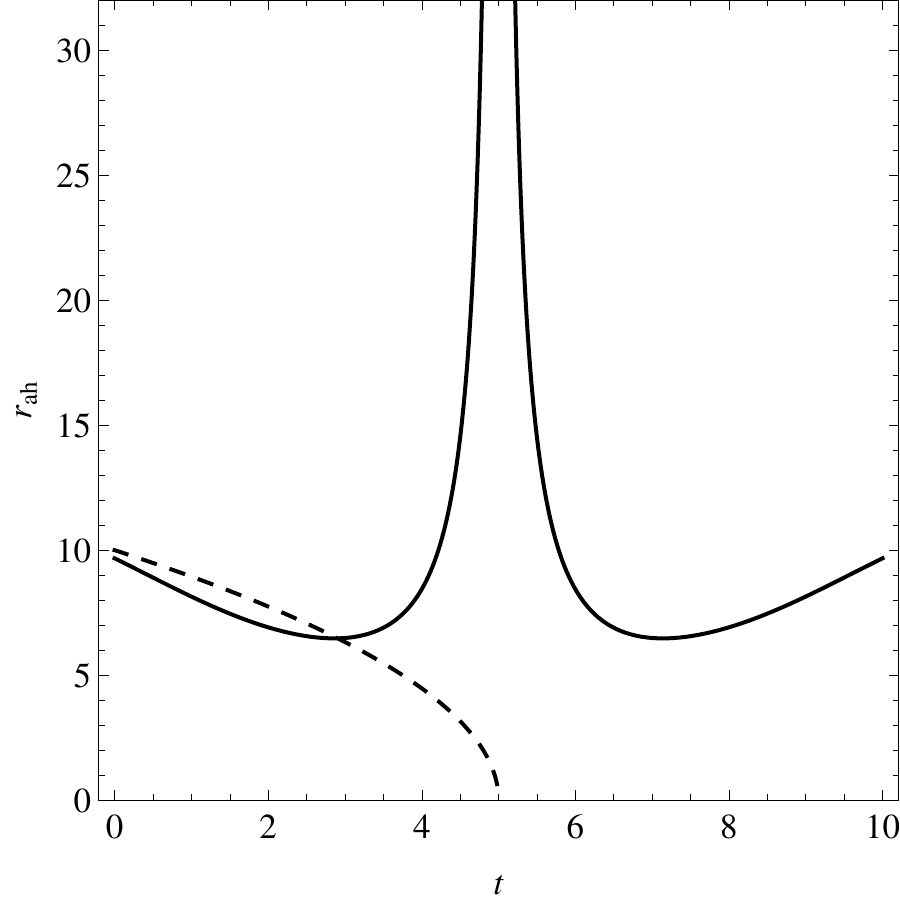}
\caption{Left panel: evolution of the scale factor $a(t)$ for the classical 
radiation collapse (dashed curve) and the semiclassical radiation collapse 
(solid curve) in the case $n=2$. Central panel: as in the left panel for the 
square of the Hubble rate $H^2 = (\dot{a}/a)^2$. Right panel: evolution of 
the radius of the apparent horizon $r_{\rm ah}$ for the classical radiation 
collapse (dashed curve) and semiclassical radiation collapse (solid curve). 
Here $t_0 = 5$ and $\Lambda = 1$} 
\label{f1app}
\end{figure*}

\section{Models with form factor $\exp (- \Box)^n$ \label{s-a-3}}

Finally, we have calculated some solutions for the entire function $V(z)^{-1} = e^{z^n}$ 
and higher even values of $n$. It turns out that all these models describe exactly 
the same physics, in the sense that they have the same qualitative behavior.
The solutions are much more complicated. The case $n=2$ has homogeneous 
solutions
\be
&& \hspace{-0.7cm}
a^2_{\rm radiation}(t) = \Bigg[
2 \Gamma \left(\frac{3}{4}\right) \,
   _1F_3\left(-\frac{1}{4};\frac{1}{4},\frac{1}{2},\frac{3}{4};\frac{t^4 \Lambda
   ^4}{256}\right) \nonumber \\
&&\hspace{-0.7cm}
   +\Lambda ^2 t^2 \Gamma \left(\frac{5}{4}\right) \,
   _1F_3\left(\frac{1}{4};\frac{3}{4},\frac{5}{4},\frac{3}{2};\frac{t^4 \Lambda
   ^4}{256}\right) \Bigg]\frac{1}{\pi  \Lambda  t_0} \, ,  \nonumber \\
 &&  \hspace{-0.7cm}
 a^2_{\rm dust}(t) = \frac{1}{3 \pi  \left(\Lambda  t_0\right){}^{\frac{4}{3}}}
   \Bigg[  2 \pi  \, _1F_3\left(-\frac{1}{3};\frac{1}{4},\frac{1}{2},\frac{3}{4};\frac{t^4
   \Lambda ^4}{256}\right) \nonumber \\
   &&\hspace{-0.7cm}
   +\sqrt{3} \Lambda ^2 t^2 \Gamma \left(\frac{1}{3}\right) \Gamma
   \left(\frac{7}{6}\right) \,
   _1F_3\left(\frac{1}{6};\frac{3}{4},\frac{5}{4},\frac{3}{2};\frac{t^4 \Lambda
   ^4}{256}\right)\Bigg]  . 
\ee
The plots of the scale factor, Hubble rate, and apparent horizon are shown
in Fig~\ref{f1app}. We have also checked the cases with $n=4$ and 12, finding
very similar plots.


\end{document}